\documentclass[preprint]{aastex701}

\definecolor{Red}{rgb}{1,0,0}

\definecolor{terry}{rgb}{0.0,0.5,0.0}

\received{} 
\revised{} 
\accepted{} 
\submitjournal{ApJ}

\shorttitle{Modeling Quasi-Periodic Fast Magnetosonic Waves}
\shortauthors{Ofman et al.}

\begin{document}

\title{Modeling the Excitation, Propagation and Damping of Quasi-Periodic Fast Magnetosonic Waves in Realistic Coronal Active Region Magnetic Field Structures}

\correspondingauthor{Leon Ofman}
\author[0000-0003-0602-6693]{Leon Ofman}
\affiliation{Department of Physics, 
Catholic University of America, 
Washington, DC 20064, USA}
\affiliation{Heliophysics Science Division, 
NASA Goddard Space Flight Center, 
Greenbelt, MD 20771, USA}
\altaffiliation{Visiting, Department of Geosciences,  Tel Aviv University, Tel Aviv, Israel}
\email{show}{ofman@cua.edu}

\author[0000-0003-0053-1146]{Tongjiang Wang}
\affiliation{Department of Physics,
Catholic University of America, 
Washington, DC 20064, USA}
\affiliation{Heliophysics Science Division, 
NASA Goddard Space Flight Center, 
Greenbelt, MD 20771, USA}
\email{wangt@cua.edu}


\author[0000-0003-4043-616X]{Xudong Sun}
\affiliation{Institute for Astronomy, 
University of Hawaii at M\={a}noa,
Honolulu, HI 96822, USA}
\email{xudongs@hawaii.edu}

\author[0000-0002-9672-3873]{Meng Jin}
\affiliation{Lockheed Martin Solar and Astrophysics Laboratory, Building 203, 3251 Hanover Street, Palo Alto, CA 94304, USA}
\email{jinmeng@lmsal.com}




\begin{abstract}
Quasi-periodic fast propagating magnetosonic waves (QFPs) were discovered in the solar corona in EUV since the launch of SDO spacecraft more than a decade ago. The QFP waves are associated with flares and coronal mass ejections (CMEs) providing information on flare pulsations as well as on the magnetic field by MHD wave seismology. Previous models of QFP waves used primarily idealized magnetic active region structures. However, more realistic active region numerical models are needed to improve the application of coronal seismology to observations of waves in coronal structures. Here, we extend the previous models by including realistic magnetic configuration based on an observed coronal active region in a case study using AR 11166 observed on March 10, 2011 by SDO/AIA, using potential field extrapolation of photospheric magnetic field with realistic gravitationally stratified density structure {  with typical coronal temperature} in our resistive 3D MHD model. We aim at reproducing the observed QFPs properties, such as directionality, propagation, reflection, nonlinearity, and damping of these waves. We model various forms of excitation of QFPs through time dependent boundary conditions, and localized pulses at the base of the corona. We produce synthetic emission measure (EM) maps from the 3D MHD modeling results to facilitate comparison to EUV observations. We find that the present more realistic model provides better qualitative agreement with observations compared to previous idealized models, improving the study of QFP wave excitation, propagation and damping in coronal ARs, with potential applications to coronal seismology.
\end{abstract}



\section{Introduction} \label{intro:sec}
High resolution and high cadence observations of coronal active regions (ARs) with Solar Dynamics Observatory \citep[SDO;][]{Pes12} Atmospheric Imaging Assemble (AIA, \citep[AIA;][]{Lem12} in Extreme Ultraviolet (EUV) emission, as well as with other spacecraft reveal ample evidence of MHD waves in coronal structures \citep[see the reviews][]{LO14,Wan16,NK20,Ban21,Wan21}.  The observed quasi-periodic {  flare pulsations (QPPs, see the review \citet{Van16})} are found to be associated with the observed quasi-periodic MHD waves { in many studies, first detected by \citet{Liu11}}, where the location and directionality, magnitude of the flare, as well as the active region (AR) magnetic geometry  determine the wave properties. The detection and analysis of MHD waves in the solar corona provide a useful tool for coronal seismology, i.e., where the unknown coronal parameters are determined from the wave proprieties \citep[see the review][]{Nak24}. The QFP waves were also observed behind global EUV waves generated by propagating CME shock fronts \citep{Liu12,She22}. 

{ In the large-scale corona the fast mode waves interact with a random ensemble of coronal loops, open field regions, magnetic nulls, and other magnetic structures. The waves may  resonate with some coronal loops causing kink oscillations and resonant energy transfer \citep[e.g.,][]{Mur01,OT02,Bal05,LO14,Guo15,Nak24}. When the propagating fast magnetosonic waves imping on magnetic singularities like magnetic nulls  then there is more complex dissipation process at play due to the interaction of waves and reconnection-generated plasma, waves, beams and associated energy dissipation \citep[see, e.g.,][and references within]{Sri25}; these complex interactions are beyond the scope of the present study.}

This study is focused on propagating quasi-period fast magnetosonic (QFP) wave trains associated with flare pulsations that were first discovered by \citet{Liu11,Liu12}, using SDO/AIA observations, and since then QFP wave observations were reported in many studies \citep[e.g.][]{She12,She13,She17,She18,She22,Yua13,Nis14,Zha15,God16,OL18,Li18a,Mia19,Mia20,Mia21,Zho22,Zho24}. The waves were analyzed, identified and studied as fast magnetosonic wave trains produced by impulsive events using idealized MHD models of coronal structures \citep{Ofm11,LO14,Nis14,TS16,Pas17,OL18,God19,Kol21,Mon24,Hu24}. { In the above numerous observational studies the potential importance of the QFP waves in energy transfer from flare sites and CME shock fronts to the extended corona became evident and the need for more realistic 3D MHD modeling of QFP wave excitation, propagation, and dissipation became apparent.}

In the present study we extend the previous models { \citep[such as][]{OT02,Ofm12,Ofm15b,OL18,OW22}} by including realistic magnetic configuration of the AR based on photospheric magnetogram of AR11166 observed on on Mar-10-2011 and using potential field Green's function extrapolation method \citep{Sak89} with gravitationally stratified density to initialize the 3D MHD simulation of a QFP wave trains. { This new more realistic approach for QFP 3D MHD modeling study not previously attempted, allows more direct and detailed modeling study of QFP wave propagation and dissipation, alleviating some of the observational limitation of plane-of-the-sky (2D) SDO/AIA observations}.  We investigate various excitation parameters of the waves, as well as the effects of flare location, and compare to the coronal observations of QFPs in EUV from SDO/AIA.  We calculate the synthetic emission measure (EM) maps from the 3D MHD model output and compare to observations. 

The paper is organized as follows: in Section~\ref{obs:sec} we provide the observational motivation for this study using SDO/AIA observation of QFPs, in Section~\ref{model:sec} we describe the 3D MHD numerical model, the initial and the boundary conditions, in Section~\ref{num:sec} we present the numerical results, and in Section~\ref{dc:sec} we present the discussion and the conclusions.

\section{Observational Motivation}
\label{obs:sec}

\begin{figure}[ht]
\centering
\includegraphics[width=1.0\linewidth]{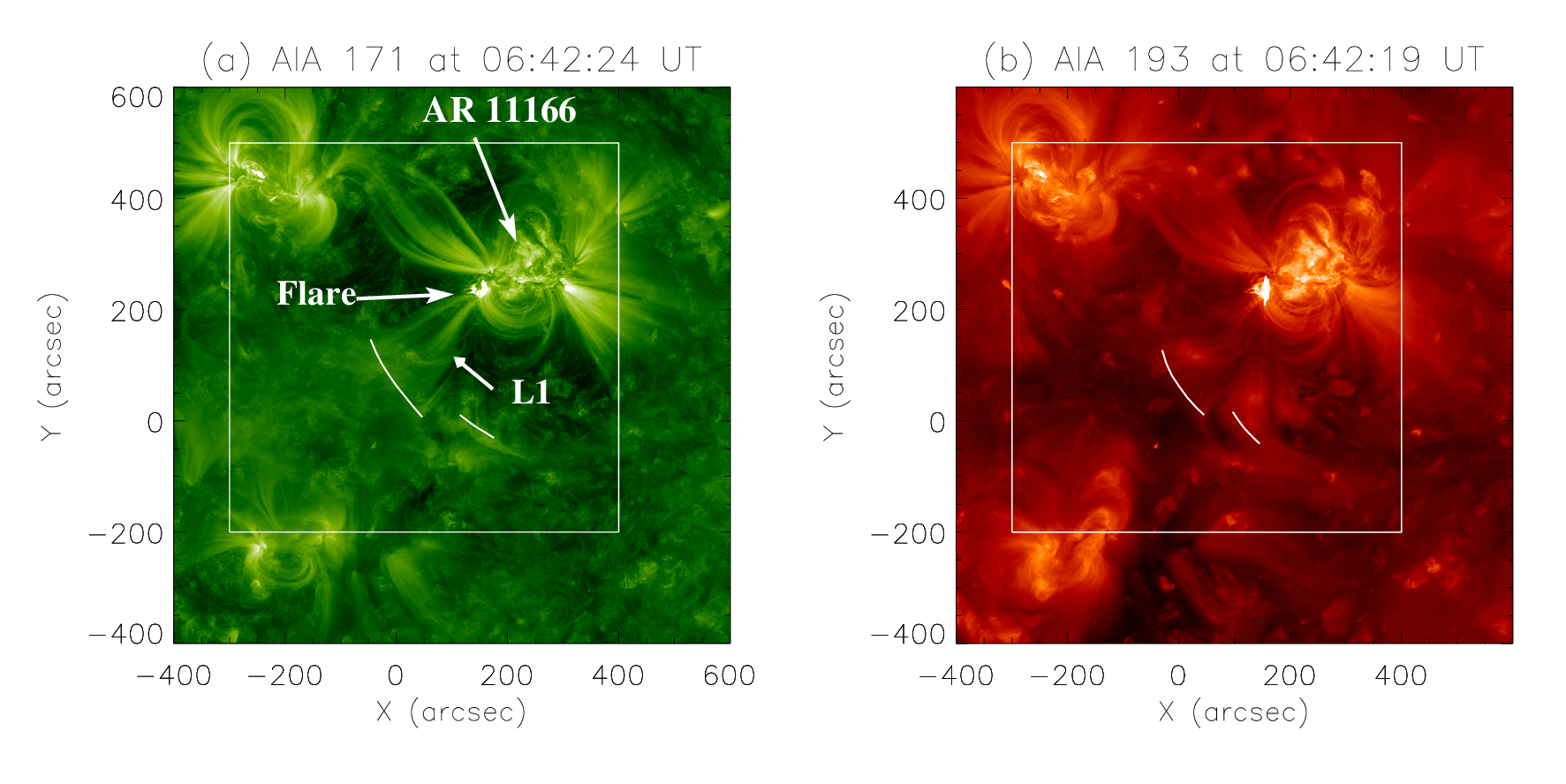}
\caption{The QFP wave event observed in AR 11166 on 2011 March 10 by SDO/AIA in (a) 171 \AA, and (b) 193 \AA~channels. White arrows in (a) mark the AR, flare source, and fan loop associated with the event. Two white curves indicate the initial EUV wave fronts. The white box marks a region, where  the propagating waves in the running difference images are shown { in} Figure~\ref{QFP_171_193_diff:fig}.  
}
\label{aiaimg:fig}
\end{figure}

We chose the SDO/AIA observations of a QFP wave train occurring in AR 11166 on Mar-10-2011 as the motivation and setup for our 3D MHD study. The QFP waves were produced by a GOES C4.0 flare in the time period about 06:40:00- 06:53:12 UT. Figure~\ref{aiaimg:fig} shows a snapshot of this wave event observed in the AIA 171 and 193 \AA\ channels. The QFP waves were analyzed in detail by \citet{Mia19}.

\begin{figure}[h]
\centering
\includegraphics[width=0.9\linewidth]{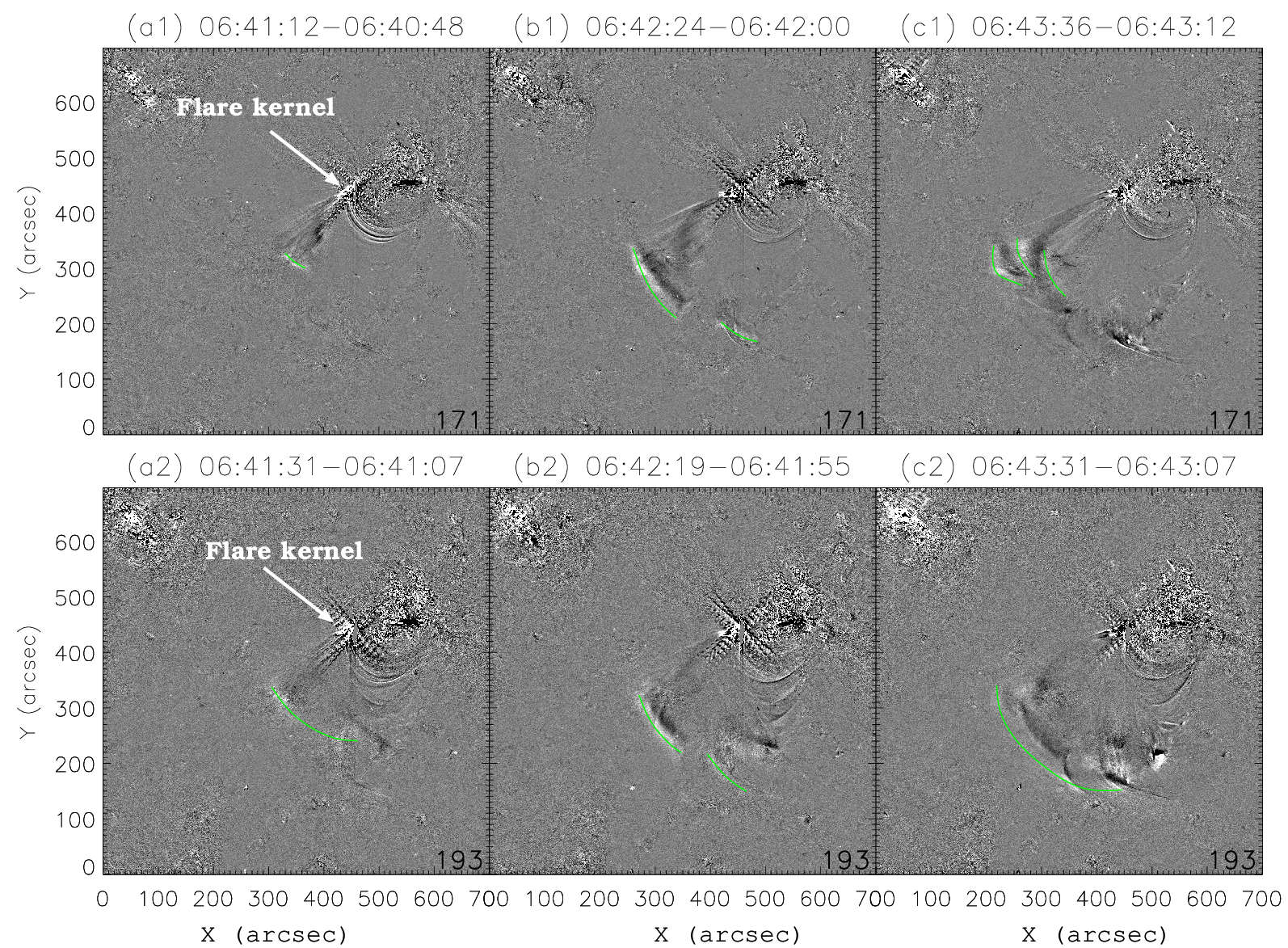}
\caption{The difference images showing the QFP waves observed by AIA in 171\AA\ (a1)$-$(c1) and 193\AA\ (a2)$-$(c2) at a sequence of times in the range 06:40$-$06:50 UT on Mar-10-2011 following the onset of the C class flare.  The green curves in each panel mark the wavefronts of EUV waves (or QFP wave trains in panel (c1)). Animations are available online. { The accelerated animation shows the time-sequence of the difference images for the two AIA EUV wavelengths of 171\AA\  (corresponding to the top panels) and the 193\AA\ (corresponding to the lower panels) for the time interval 06:40$-$06:50 UT on Mar-10-2011.}} 
\label{QFP_171_193_diff:fig}
\end{figure}

In Figure~\ref{QFP_171_193_diff:fig},  the QFP wave train is evident in the running-difference images obtained from  AIA in 171\AA\ (panels a1-c1) and 193\AA\ (panels a2-c2) observations at a sequence of times in the range 06:40-06:50 UT on Mar-10-2011 (animations are available online). The QFP waves appear following the onset of the C class flare at the location in the upper right corner of the panels, evident in the brighter saturated emission.  The wave fronts of the propagating QFP wave trains are indicated with green curves in the various panels. 

\begin{figure}[h]
\centering
\includegraphics[width=0.5\linewidth]{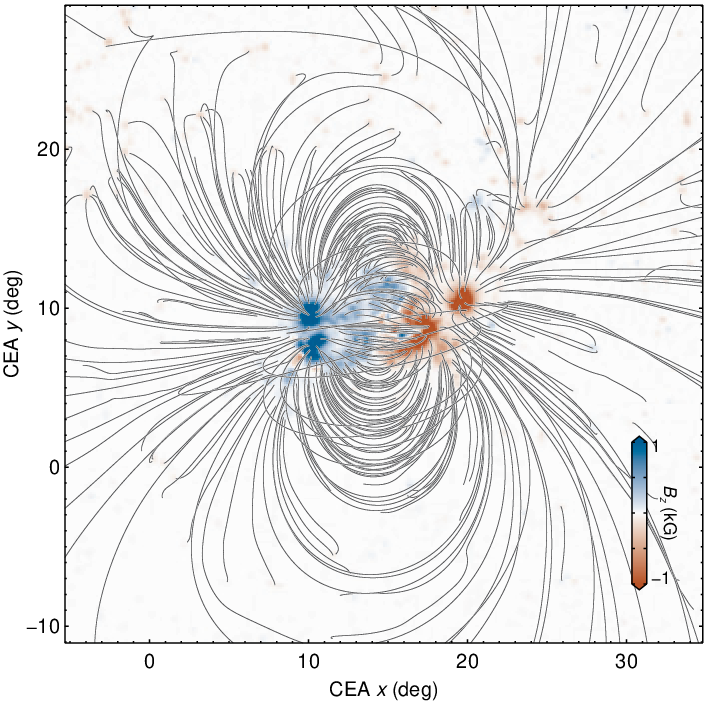}
\caption{The potential magnetic field extrapolation using the HMI vector magnetogram of AR 11166, the source of the QFP waves observed by SDO/AIA on Mar-10-2011 at about 06:40 UT. The red (positive polarity) and blue (negative polarity) colors indicate the photospheric magnetic field in the range $\pm$1000 G. The ‘cylindrical equal area’ map projection (CEA) coordinates are used here \citep[see,][]{Sun13}}
\label{AR11166:fig}
\end{figure}

The radial magnetic field, de-projected from the HMI vector magnetogram, was used to reconstruct the coronal magnetic field of this AR by using potential field with the Green's function extrapolation method. The resulting magnetic field configuration is shown in Figure~\ref{AR11166:fig}.

 The main characteristics of the 2011-03-10 wave event that motivate our modeling study are summarized below, expanding on the analysis of this event in \citet{Mia19}:
\begin{enumerate}
  \item The initial EUV wave fronts spread over a wider range than the QFP wave trains, which propagate along the fan loop system marked with L1 in Figure~\ref{aiaimg:fig}.                   

 \item The QFP waves first emerged about 36 Mm away from the flare kernel, persisting with a lifetime of 2 minutes. The QFP wave train composed of 4 wavefronts propagates with speeds of about 680$-$840 km~s$^{-1}$, approximately consistent with the main EUV wavefronts (470$-$920 km~s$^{-1}$).

 \item The QFP waves have a period of about 45 s, while the similar periodicity is not found in the light curve of flare kernels. Thus, \citet{Mia19} suggested that the QFP wave trains observed in this event may be excited by interaction of the CME shock wave with funnel-like coronal loops. During this process, the initial broadband pulse dispersively evolves into multiple QFP wavefronts (in the present study we use periodic velocity pulse given below to model the excitations QFP waves). 

 \item The energy flux of the QFP waves can be estimated using the approach provided in \citet{Liu11} and \citet{OL18} based on WKB approximation using the average observed QFP wave propagation speed of 760  km~s$^{-1}$ the typical { emission} intensity perturbation due to the wave $\delta I/I=3$\%. { The number density for the wave energy flux  estimate is obtained from the differential emission measure (DEM) analysis \citep{Che15}. From the SDO/AIA EUV observation of the wavefront region on Mar-10-2011 at 6:42 UT  (see, Figure~\ref{QFP_171_193_diff:fig}, panels b1 and b2)  we get ${ n=4.6\pm1\times10^8}$ cm$^{-3}$  with estimated LOS depth=17 Mm. The obtained density value is consistent with quite sun (QS) electron densities \citep[e.g.,][]{Der20}. U}sing Equation~(1) in \citet{OL18} we get an energy flux $E\gtrsim {(0.36-0.56)}\times10^5$ erg/cm$^2$/s comparable to coronal heating energy flux requirement \citep{WN77}. 
\end{enumerate}                                                            

\section{Numerical 3D MHD Model, Boundary and Initial Conditions} \label{model:sec}


In order to model the QFP waves, we employ the 3D MHD code NLRAT  described in previous studies of MHD waves in coronal ARs  \citep{OT02,Ofm15b,POW18,OL18,OW22,wang24}. The code solves the resistive 3D MHD equations with gravity, and  standard notation for the variables, written as 
\begin{eqnarray}
&&\frac{\partial\rho}{\partial t}+\nabla\cdot(\rho\mbox{\bf V})=0,\label{cont:eq}\\
&&\frac{\partial(\rho\mbox{\bf V})}{\partial t}+\nabla\cdot\left[\rho\mbox{\bf V}\mbox{\bf V}+\left(E_up+\frac{\mbox{\bf B}\cdot \mbox{\bf B}}{2}\right)\mbox{\bf I}-\mbox{\bf BB}\right]=-\frac{1}{F_r}\rho\mbox{\bf F}_g,\label{mom:eq}\\
&&\frac{\partial\mbox{\bf B}}{\partial t}=\nabla\times(\mbox{\bf V}\times\mbox{\bf B})+\frac{1}{S}\nabla^2\mbox{\bf B},\label{ind:eq}\\
&&\frac{\partial(\rho E)}{\partial t}+\nabla\cdot\left[\mbox{\bf V}\left(\rho E+E_up+\frac{\mbox{\bf B}\cdot \mbox{\bf B}}{2}\right)-\mbox{\bf B}(\mbox{\bf B}\cdot\mbox{\bf V})+\frac{1}{S}\nabla\times\mbox{\bf B}\times\mbox{\bf B}\right] 
=-\frac{1}{F_r}\rho\mbox{\bf F}_g\cdot\mbox{\bf V}
,\label{ener:eq} 
\end{eqnarray}
where $\mbox{\bf F}_g=\frac{L_0^2}{(R_s+\Delta z)^2}\mbox{\bf\^z}$ is the gravitational force modeled with the assumption of small Cartesian box in the corona place at solar radius $R_s$, where $L_0=0.1R_s$, $\Delta z=z-z_0$ is the height of the lower boundary of the model AR,  $\rho E=\frac{E_up}{(\gamma-1)}+\frac{\rho V^2}{2}+\frac{B^2}{2}$ is the total energy density. Further details of the 3D MHD model and the normalizations can be found in \citet{OL18} and references therein. In the present study we use the normalization parameters $B_0=244$ G, $n_0=2\times10^9$ cm$^{-3}$, $T_0=10^6$ K. In the present model {the Lundquist number is set to $S=10^4$ as in previous models, resulting in enhanced (compared to coronal values) but still small resistive dissipation. Here,} we neglect thermal conduction, viscosity, and radiative cooling, since these terms  have small effect on the properties of fast magnetosonic QFP waves and the short timescale of the event. 

\begin{figure}[h]
\centering
\includegraphics[width=0.8\linewidth]{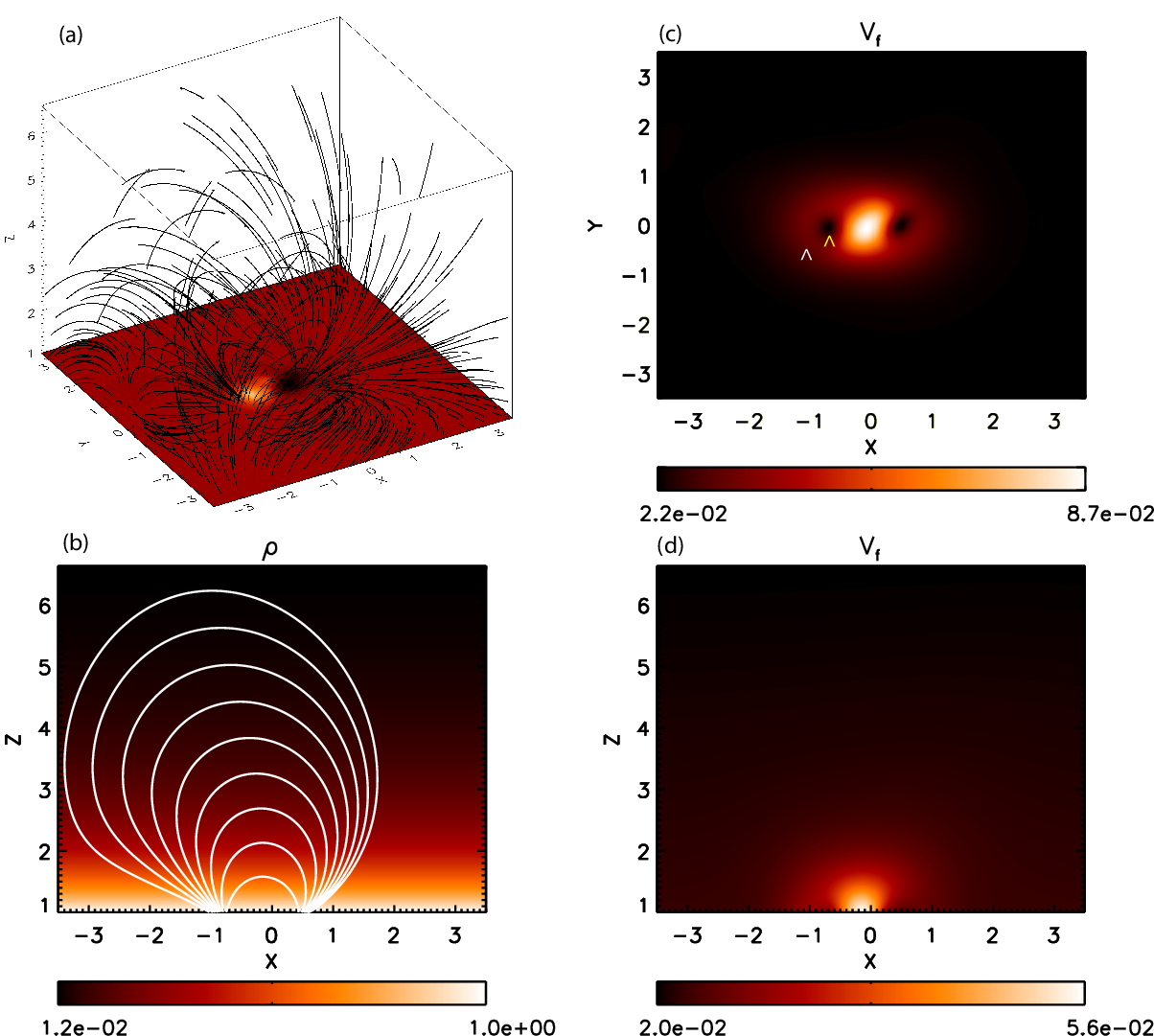}
\caption{The initial state of the 3D MHD model $(t=0)$. (a) The magnetic field lines reconstructed using potential field extrapolation with Green's function method of the radial magnetic field for AR 11166. (b) The gravitationally stratified density in the $x-z$ plane at $y={ -0.72}$, overlaid with several fieldlines that are calculated from the ($B_x$, $B_z$) components in this 2D plane. (c) The fast magnetosonic speed, $V_f$, in the $x-y$ plane at ${ z=1.42}$. The arrowheads mark the location of the QFP wave train source in the 3D MHD model: Source A at $(-0.71, -0.2)$ (yellow arrowhead), and Source B at $(-1.1, -0.5)$ (white arrowhead). (d) $V_f$ in the $x-z$ plane at ${ y=-0.72}$.} 
\label{initBnVf:fig}
\end{figure}

The initial state of the 3D MHD model is shown in Figure~\ref{initBnVf:fig}, where the initial magnetic field configuration was reconstructed using the potential field extrapolation from the radial component of the vector magnetic fields for AR 11166. In the simulation domain, the coronal lower boundary is placed at 12.37 Mm above the photosphere with a maximum magnetic field magnitude of 244 Gauss for this AR. The reason for this approach is to avoid modeling the very high magnetic field magnitude region in the lower part of the corona, where the Alfv\'{e}n speed would be extremely high, reducing the time step to an unpractical low value. This does not affect the results considerably, since the QFP wave propagates mostly upward along the gradient of the fast magnetosonic speed.

Figure~\ref{initBnVf:fig}a shows the reconstructed magnetic field lines in the initial state. The initial gravitationally stratified density in the $x-z$ plane cut at $y=0$ is shown in Figure~\ref{initBnVf:fig}b, with several fieldlines indicated with white lines. The normalized fast magnetosonic speed for perpendicular propagation, $V_f=\sqrt{V_A^2+C_s^2}$, in $x-y$ plane is shown in Figure~\ref{initBnVf:fig}c and that in the $x-z$ plane is shown in Figure~\ref{initBnVf:fig}d, indicating the fastest $V_f$ in the lower center of the closed field region of the AR, where the magnetic field strength is maximal. The velocity is normalized in units of the  Alfv\'{e}n speed, set as $V_{A,0}=B_0/(4\pi \rho_0)^{1/2}=11900$ km s$^{-1}$ based on the above $B_0$ and $n_0$ parameters, where $\rho_0=m_pn_0$ is the corresponding mass density normalization. The sound speed is $C_s=132$ km/s with $T_0=10^6$ K and $\gamma=1.05$. The Alfv\'{e}n time is defined as $\tau_A=0.1R_s/V_{A,0}=5.9$ s.   
The typical resolution of the 3D MHD model is $318\times318\times258$ grid cells with the grid size of 0.0021875$R_s$, equal to about 2.18 Mm.

The AR is modeled by solving the above 3D MHD equations, and is  first run to a steady state by evolving the model until no significant changes in the magnetic field, density, and thermal structure were detected ($t_0\approx 39\tau_A$). There is minor difference between the steady state magnetic field and the initial potential vacuum field in the low-$\beta$ model AR (where at the base of the model AR $\beta=2.3\times10^{-4}$). Next, the QFP waves were launched by a periodic velocity pulse at the coronal base of the model AR 11166 at $z=z_0=1$, $x=x_0$, $y=y_0$, where two cases with sources at $(x_0,y_0)=(-0.71, -0.2)$ (Source A), and  $(x_0,y_0)=(-1.1, -0.5)$ (Source B) were considered (see, Figure~\ref{initBnVf:fig}c and the numerical results in Section~\ref{num:sec}). Source A is closer to the center of the model AR, while Source B is in the observed AR propagation region of the QFP waves formed by the 3D structure of the fast magnetosonic speed, $V_f$. The structure of $V_f$ affects the directionality, spatial divergence, and slowing/damping of the waves similar to a weveguide for fast MHD waves \citep[e.g.][]{OD95,Pas13,Pas14}. Although in the present case there are no sharp waveguide boundaries, and therefore the refraction of the waves is small and the dispersive effects are negligible. 

The following form of the velocity perturbation due to the flare pulsations was applied after a steady state AR model was obtained (typically) at  $t\gtrsim 36\tau_A$:
\begin{eqnarray}
&&V_x(x,y,z=z_0,t)=V_d\sin(\omega t)\,e^{-[(x-x_0)/w]^2-[(y-y_0)/w]^2},
\label{pulse:eq} 
\end{eqnarray} where the amplitude of the source was set to $V_d=0.02V_{A,0}$, the normalized frequency $\omega=0.249$ corresponding to a period of 148 s ($\sim$2.5 min), consistent with  typical range of QFP wave periods observed in the corona \citep[e.g., see the reviews][]{LO14,She22}. 
The spatial width of the QFP wave source is $w=0.125$ (corresponding to $0.0125R_s$), which is small compared to the size of the AR, modeling qualitatively the localized nature of the { wave source}. The functional form of the velocity perturbation is similar to our previous studies \citep[e.g.,][]{OL18}, where the $x$-component of velocity was driven at the boundary, other velocity and magnetic field components are driven self-consistently due to the coupling by incoming characteristic to the boundary, approximated by extrapolated variables in the 3D MHD model equations \citep[e.g.,][]{OT02}.

\section{Numerical Results} \label{num:sec}

\begin{figure}[h]
\centering
\includegraphics[width=0.8\linewidth]{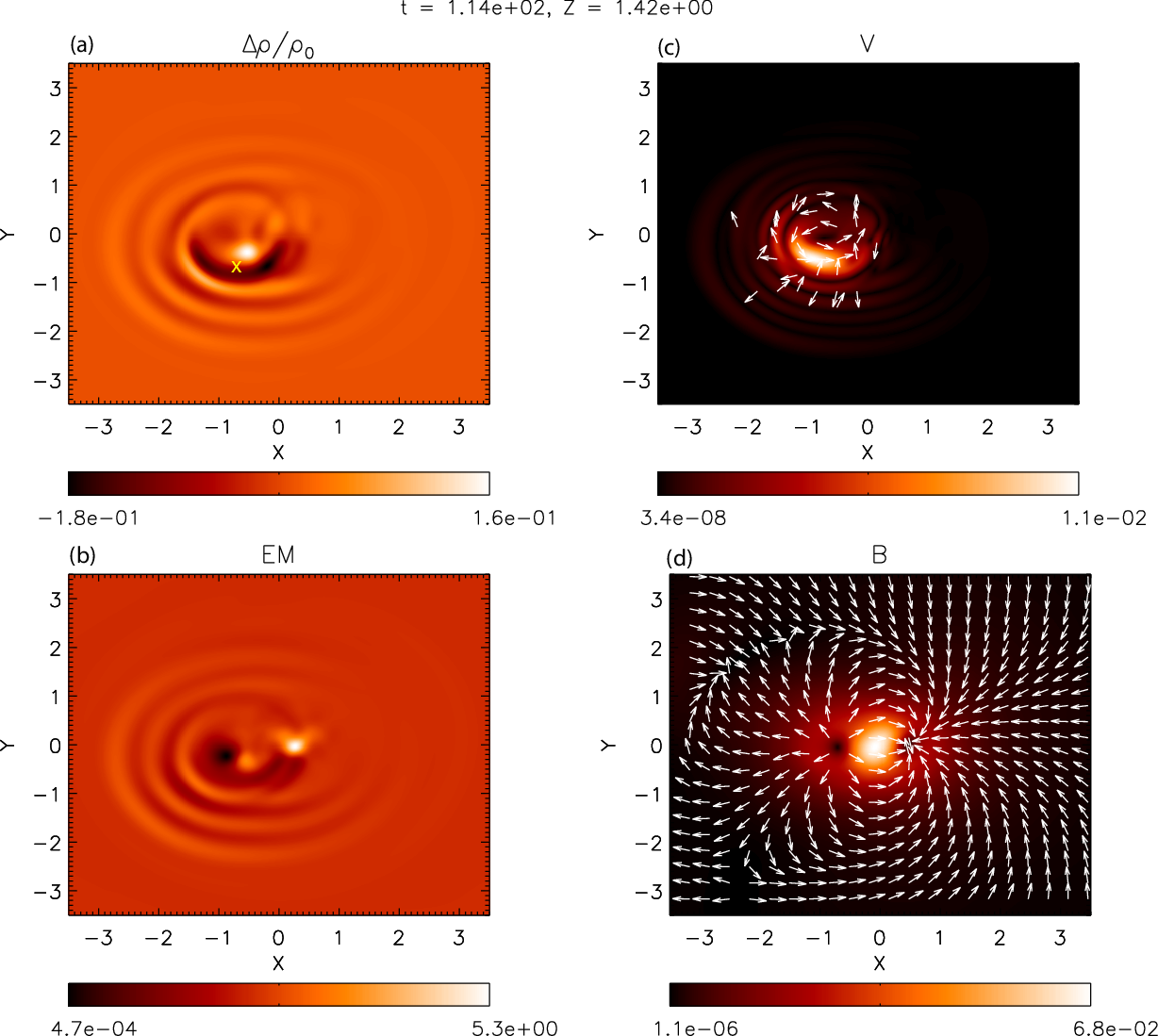}
\caption{The snapshot of the variables in the $x-y$ plane  (`on disk') at $t=114\tau_A$, $z=1.42$ due to the  QFP waves  launched at Source A in the modeled AR 11166. (a) The relative 'running difference' density perturbation $\Delta\rho/\rho_0$, where $\Delta\rho=(\rho(t)-\rho(t-\Delta t)/\rho_0$; the yellow `x' marks the location of the temporal evolution shown below. (b) the running difference of the emission measure, EM, computed from the 3D model AR. { The online accelerated animation of this panel shows the modeled time interval $t=[42.1, 129]\tau_A$ (corresponding to $\sim$ 8:33 min. duration).}  (c) The velocity magnitude and direction. (d) The magnetic field magnitude and direction. }
\label{drhVEMB_xyA_t114:fig}
\end{figure}

\begin{figure}[h]
\centering
\includegraphics[width=0.8\linewidth]{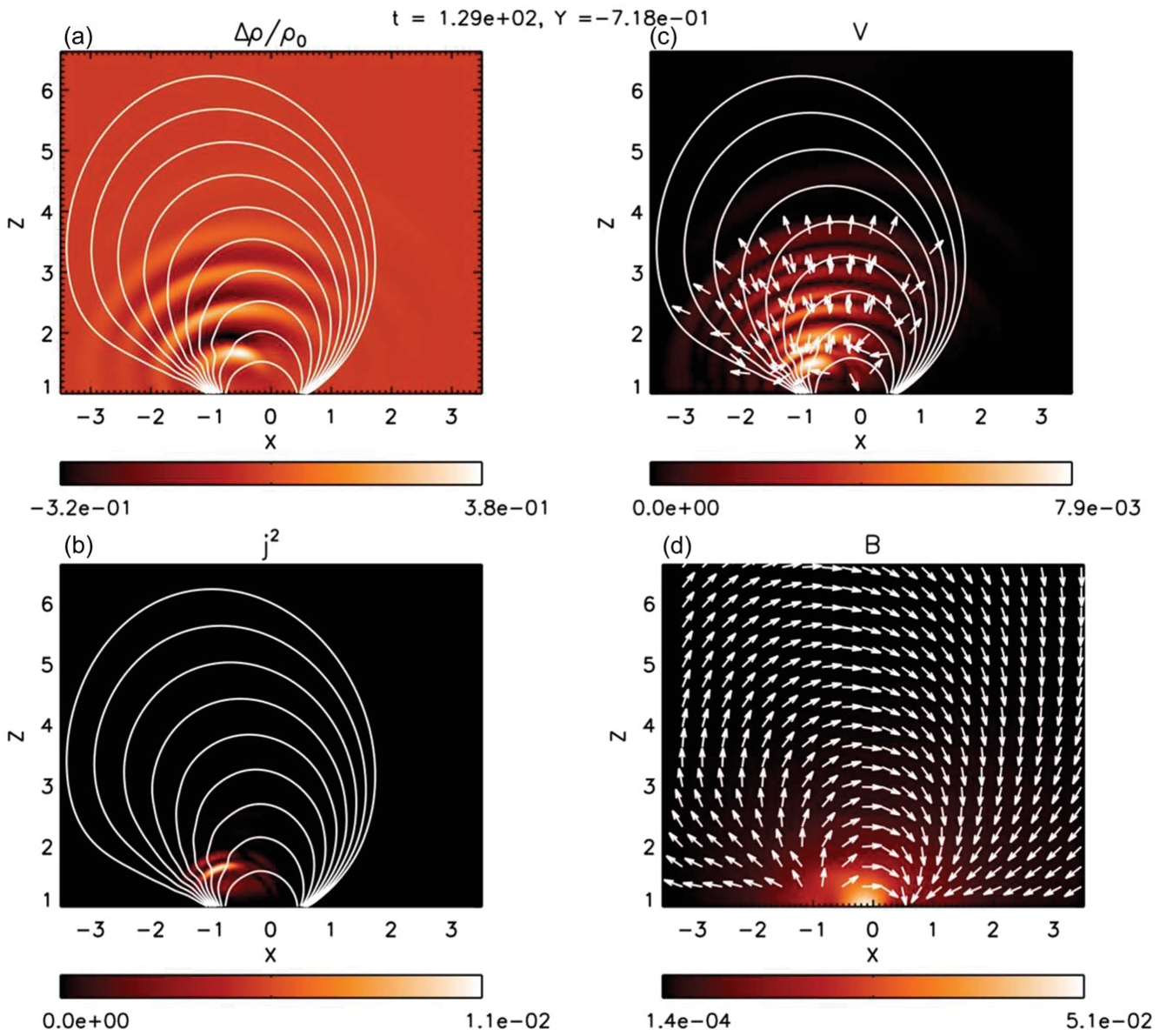}
\caption{The snapshot of the variables in the $x-z$ (`off limb') plane at $t=129\tau_A$ due to the  QFP waves launched at Source A in the modeled AR 11166. (a) The relative density perturbation $\Delta\rho/\rho_0$ with several overlaid magnetic fieldlines computed from the ($B_x,B_z$) components in the $x-z$ plane at $y=-0.72$. { The online accelerated animation of this panel shows the modeled time interval $t=[42.1, 129]\tau_A$ (corresponding $\sim$ 8:33 min. duration).} 
(b) the current density magnitude squared, $j^2$. (c) The velocity magnitude and direction arrows with several overlaid magnetic fieldlines. (d) The magnetic field magnitude and direction arrows. }
\label{drhVEMB_xzA_t129:fig}
\end{figure}

\begin{figure}[h]
\centering
\includegraphics[width=0.5\linewidth]{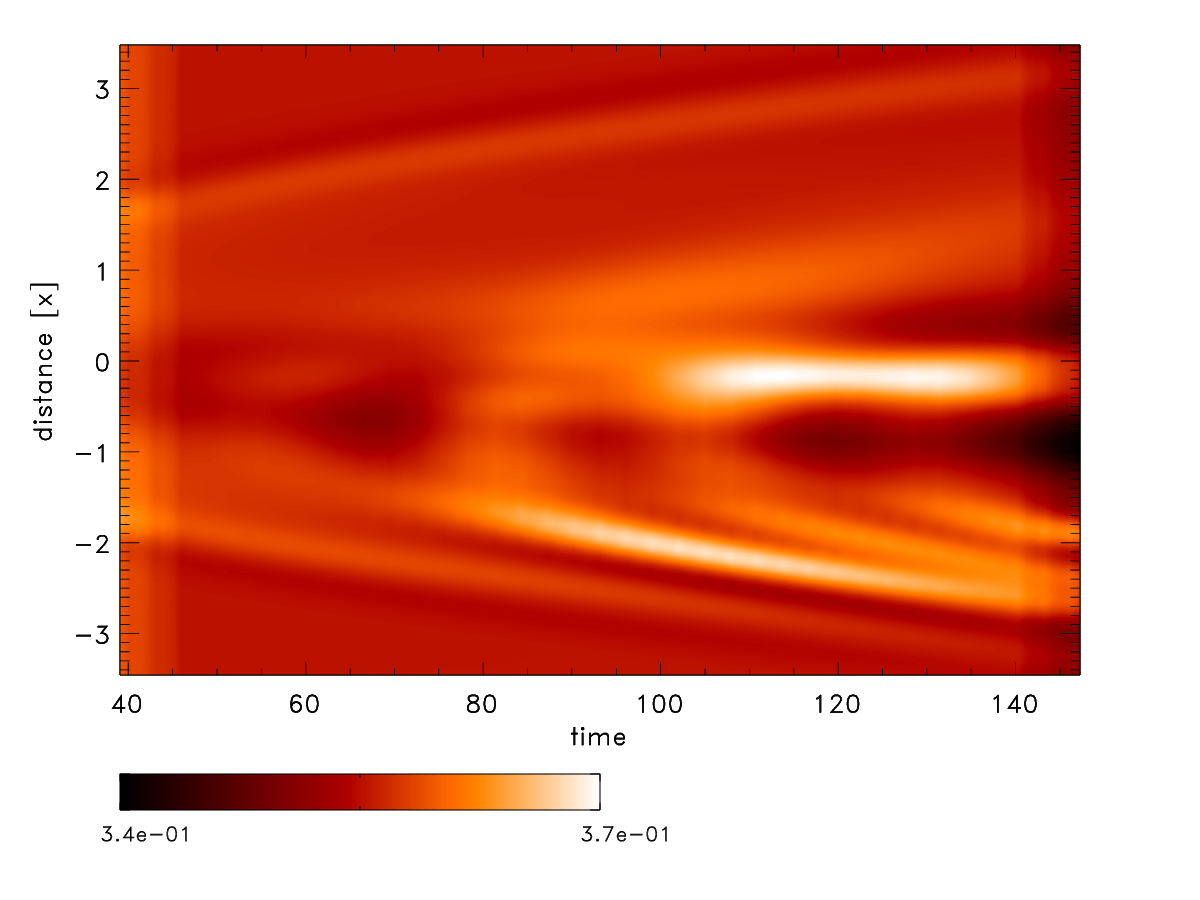}
\caption{ The time-distance plot of the perturbed density obtained from the 3D MHD model of AR11166 for the case in Figure~\ref{drhVEMB_xzA_t129:fig}a at height $z=2$, at $y=-0.72$.  The temporal cadence is 3$\tau_A$ in this figure. 
 }
\label{tdx_z2:fig}
\end{figure}

\begin{figure}[h]
\centering
\includegraphics[width=0.5\linewidth]{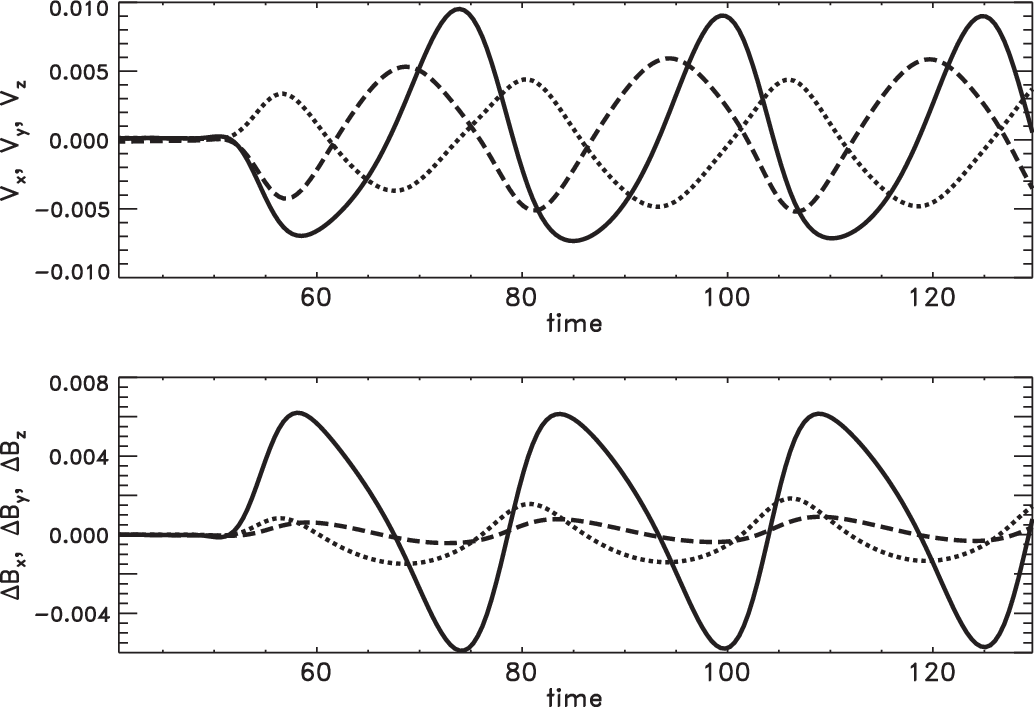}
\caption{The temporal evolution of the velocity components (top panel), and the magnetic field components perturbations $\Delta B_{x,y,z}$ (lower panel) at the location $(-0.7, -0.7, 1.7)$ for QFP waves with the parameters $V_d=0.02$, $\omega=0.249$  launched at Source A in the modeled AR 11166. The start time of the plot corresponds to the launch time of the perturbation. The sampled spatial location  in $x-y$ plane is marked with yellow `x' in Figure~\ref{drhVEMB_xyA_t114:fig}a. The line-styles are { solid}: $x$-components, { dashes}: $y$-components, { dots}: $z$-components. The time is in units of $\tau_A=5.9$ s, the total time corresponds to 12.74 min.}
\label{vdbtA:fig}
\end{figure}

\begin{figure}[h]
\centering
\includegraphics[width=0.8\linewidth]{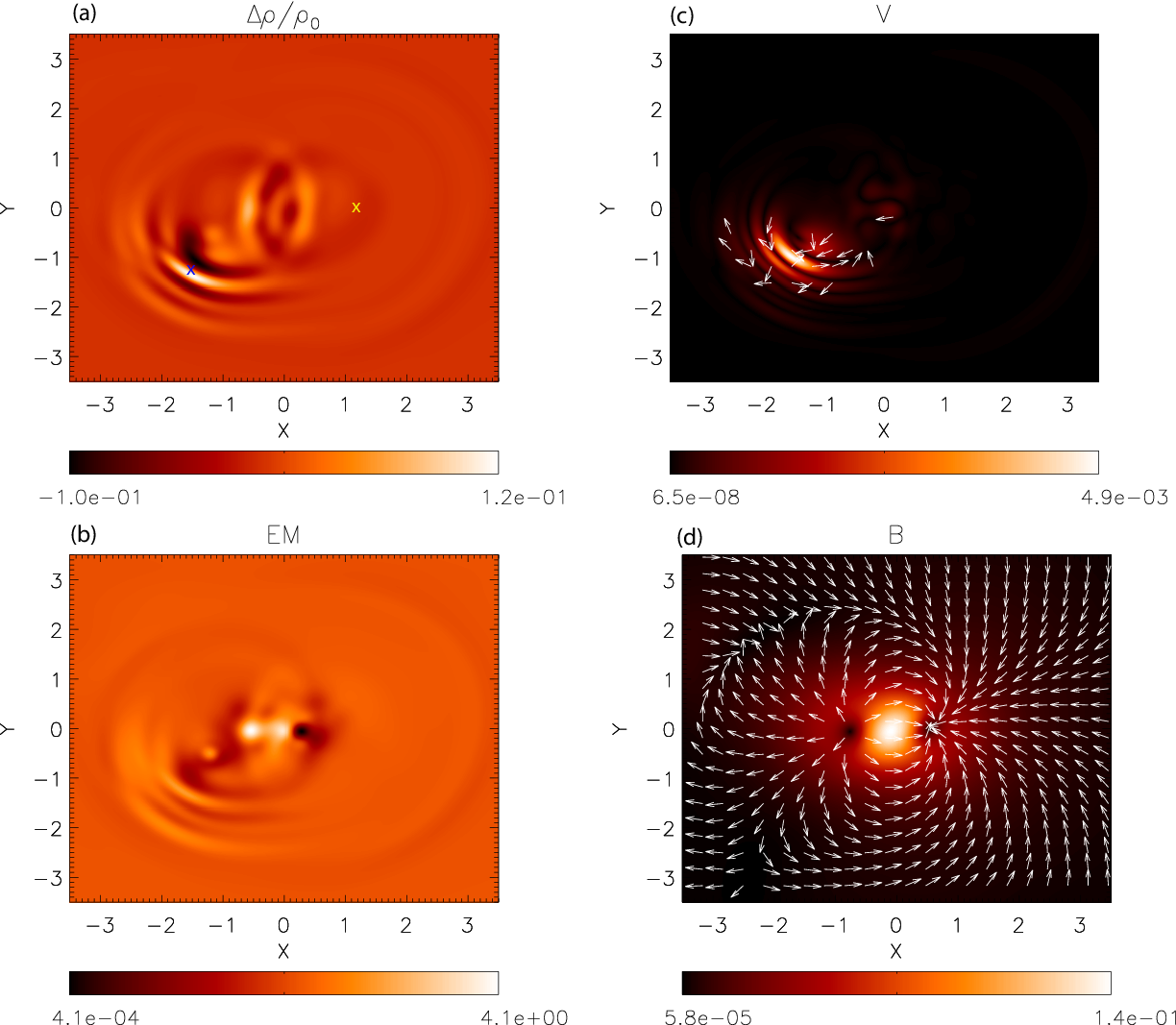}
\caption{The snapshot of the variables in the $x-y$ plane  (`on disk') at $t=123\tau_A$, $z=1.42$ due to the  QFP waves launched at Source B in the modeled AR 11166. (a) The relative density perturbation $\Delta\rho/\rho_0$, the blue and the yellow `x' mark the location of the temporal evolution shown below. (b) the emission measure computed from the 3D model AR.  (c) The velocity magnitude and direction. (d) The magnetic field magnitude and direction. { The animation of the running difference of the density and EM is available on line. The accelerated animation of the density perturbation (i.e., running difference) is shown for the modeled time interval $t=[42.1,99.2]\tau_A$ ($\sim$5:37 min duration) and for the EM  at time interval $t=[42.1,147]\tau_A$ ($\sim$10:19 min duration).}}
\label{drhVEMB_xyB_t123:fig}
\end{figure}

\begin{figure}[h]
\centering
\includegraphics[width=0.8\linewidth]{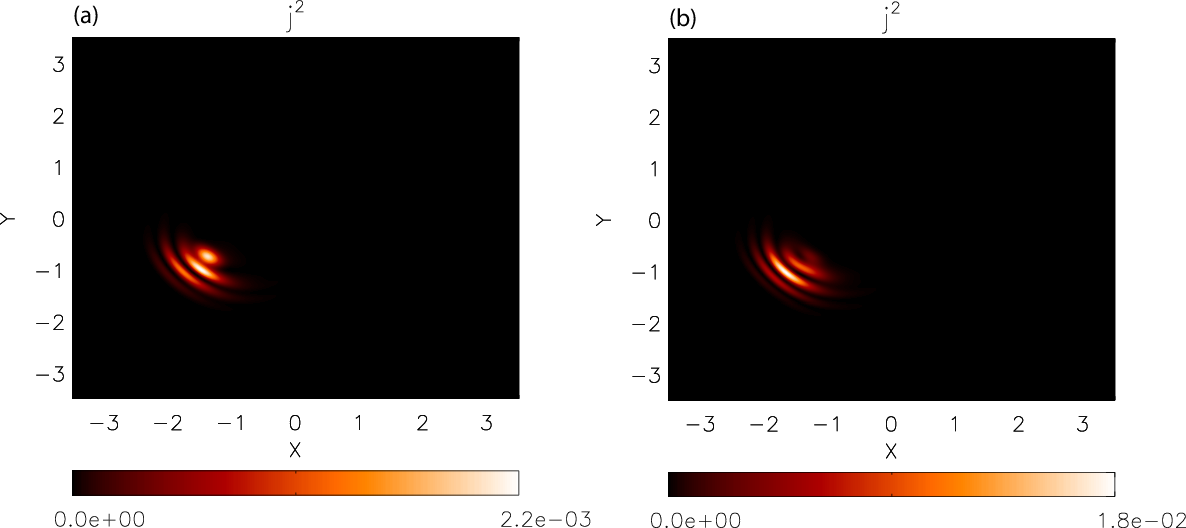}
\caption{The current density squared $j^2$ in the $x-y$ plane  (`on disk') at $z=1.42$ due to the  QFP waves launched at Source B in the modeled AR 11166  at times (a) $t=78\tau_A$, and (b) $t=123\tau_A$. Animation of this figure is available online. { The accelerated animation show the temporal evolution of $j^2$ in the  $x-y$ plane  (`on disk') at height $z=1.42$ for the modeling time interval 39-147$\tau_A$ (corresponding to $\sim$10:37 min. duration).} }
\label{j2_xyB_t78-123:fig}
\end{figure}

\begin{figure}[h]
(a)\hspace{4in}(b)\\
\includegraphics[width=1.0\linewidth]{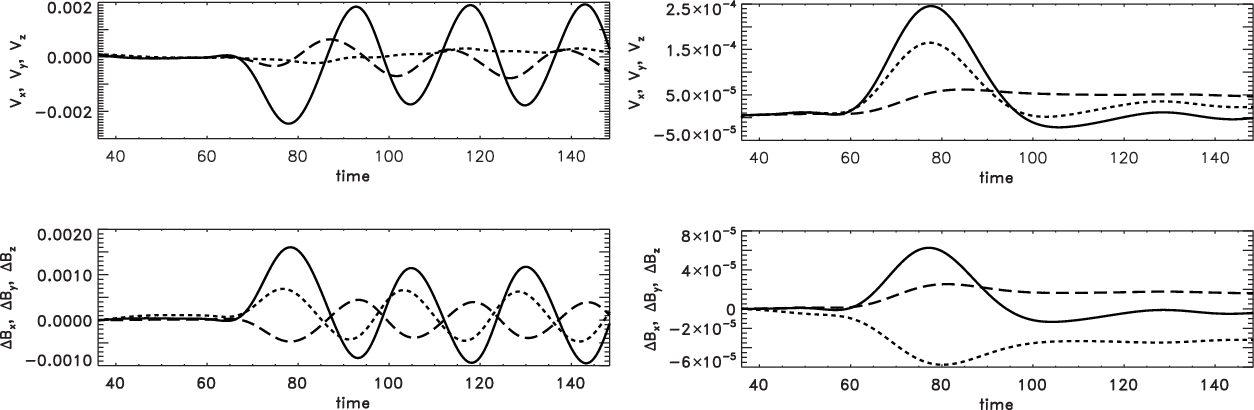}
\caption{The temporal evolution of the velocity components and perturbed magnetic field components, $\Delta B_{x,y,z}$ due to the  QFP waves launched at Source B in the modeled AR 11166. (a) The temporal evolution at  the location $(-1.49,-1.{36},1.7)$ marked with blue `x' in Figure~\ref{drhVEMB_xyB_t123:fig}a; (b) the temporal evolution for the same case but at location $(1.20,-0.01,3.5)$ marked with yellow `x' in Figure~\ref{drhVEMB_xyB_t123:fig}a. The line-styles are { solid}: $x$-components, { dashes}: $y$-components, { dots}: $z$-components.}
\label{dbvnTt_p2:fig}
\end{figure}

In Figures~\ref{drhVEMB_xyA_t114:fig}-\ref{dbvnTt_p2:fig}, we present the results of the 3D MHD modeling of the QFPs in AR 11166 for two cases of QFP wave train sources initiated at locations: Source A at $(x_0,y_0)=(-0.71, -0.2)$ and  Source B at $(x_0,y_0)=(-1.1, -0.5)$. The location of the two sources were chosen to demonstrate the effects of the AR magnetic structure on the waves. Since in the present modeling study, we focus on the source region of QFPs in AR 11166, the nearby ARs and the reflected wave effects occurring at later times of the event are not modeled. The model implements the flare pulsation scenario as the driver of the waves, while other QFP wave excitation mechanism associated with CME shock front propagation are possible (see the discussion in \citet{Liu12,LO14,Mia19}).

The results of the 3D MHD modeling of QFP wave train launched in the model of AR 11166 at Source A are shown in Figures~\ref{drhVEMB_xyA_t114:fig}-\ref{vdbtA:fig}. The variables in $x-y$ plane at height $z=1.42$ are shown in Figure~\ref{drhVEMB_xyA_t114:fig} at time $t=114\tau_A$, as the waves propagated away from the source. The Figure~\ref{drhVEMB_xyA_t114:fig}a shows the relative density perturbation $\Delta\rho/\rho_0$ due to the  QFP waves. The directionality of the wave propagation is not evident { in this case}, with nearly circular wavefronts, with largest magnitude at the source.  The normalized emission measure, EM, was computed from the 3D model AR 11166 by integrating the density square over the `line of sight' where it is assumed that the viewing angle is along the $z$ direction, and the running difference of EM is shown in panel \ref{drhVEMB_xyA_t114:fig}b. The $x-y$ plane velocity magnitude and direction arrows due to the QFP waves are shown in Figure~\ref{drhVEMB_xyA_t114:fig}c, and the magnetic field magnitude and direction are shown in Figure~\ref{drhVEMB_xyA_t114:fig}d. The  magnetic fields magnitude peaks  at the central part of the AR, and shows evidence of the QFP wave fronts in the weak field region, where the direction of the arrows changes with respect to the background field.

In Figure~\ref{drhVEMB_xzA_t129:fig}  the results are shown in the $x-z$ (`off limb') plane at $y=-0.72$ and time $t=129\tau_A$ produced by the  QFP waves launched at Source A in the modeled AR 11166. The relative density perturbation $\Delta\rho/\rho_0$ with several overlaid magnetic field lines `off-limb' are shown. It is evident that the QFP wave train propagates away from the source in height ($z$-direction) in shell-like structures that result in quasi-periodic perturbation of the density and the magnetic field (see, Figure~\ref{drhVEMB_xzA_t129:fig}a and the animation).  The corresponding current density magnitude squared, $j^2$ (proportional to Ohmic heating),  is shown in Figure~\ref{drhVEMB_xzA_t129:fig}b, peaking as expected at the largest magnetic field line perturbation.  The associated velocity magnitude and direction arrows due to the QFP wave trains with several overlaid magnetic field lines and the magnetic field magnitude and direction arrows are shown in panels Figure~\ref{drhVEMB_xzA_t129:fig}c and d, respectively.

{ In Figure~\ref{tdx_z2:fig} we show the time-distance plot at the height $z=2$, at $y=-0.72$ for the modeled AR 11166. We use the perturbed density time sequence obtained from the 3D MHD model for the case in Figure~\ref{drhVEMB_xzA_t129:fig}a. The signatures of propagating QFP waves are clearly evident as the bright slanted streaks, while the transverse (kink) oscillations are evident as alternating bright and dark patches near $x\approx -1$ in the time-distance image. The brightening and darkening of the density are due to the compressibility of the fast magnetosonic waves (both propagating, and standing).}

The temporal evolution of the velocity components $V_x$, $V_y$, $V_z$ and the magnetic field components perturbations $\Delta B(t)_{x,y,z}=B(t)_{x,y,z}-B(t=t_0)_{x,y,z}$ at the location $(x_0,y_0,z_0)=(-0.7, -0.7, 1.7)$ are shown in Figure~\ref{vdbtA:fig} for QFP wave train, launched by flow pulses with amplitude $V_d=0.02$, and frequency $\omega=0.249$ at Source A in the modeled AR 
The lead time of $\sim 22\tau_A$ is the propagation time of the disturbance from Source~A to the `observing' point at $(-0.7, -0.7, 1.7)$. 
It is evident that the $x$ component of the velocity and the magnetic field are in anti-phase, while the $z$ components are in phase as expected for propagating fast magnetosonic waves. The $B_y$ and $V_y$ appear to be phase shifted by about quarter period, indicative of the Alfv\'{e}nic wave perturbation. The effects of nonlinearity are particularly evident in the non-sinusoidal steepened structure of $\Delta B_x$ and $V_x$.

The results of the 3D MHD modeling of QFP wave train launched in model AR 11166 at Source B are shown in Figures~\ref{drhVEMB_xyB_t123:fig}-\ref{dbvnTt_p2:fig}.  Figure~\ref{drhVEMB_xyB_t123:fig} shows the snapshot of the variables in the $x-y$ plane (`on disk') at $t=123\tau_A$, height $z=1.42$ due to the  QFP waves, where  Figure~\ref{drhVEMB_xyB_t123:fig}a shows the relative density perturbation $\Delta\rho/\rho_0$, and the running difference of the emission measure, EM, computed from the 3D model AR is shown Figure~\ref{drhVEMB_xyB_t123:fig}b. The velocity magnitude and direction arrows due to the QFP waves are shown  in Figure~\ref{drhVEMB_xyB_t123:fig}c, while the magnetic field magnitude and direction are shown in Figure~\ref{drhVEMB_xyB_t123:fig}d. The directionality and the localization of the QFP wave { propagation are clearly} evident in { the present case}, where the density perturbations, EM, and velocity perturbations are directional and follow a funnel formed by the magnetic field and density structure, i.e., the background fast magnetosonic speed (see the related animations). Here, as in the previous case the  magnetic fields magnitude peaks  at the central part of the AR, and shows evidence of the QFP wave fronts in the weak field region, where the direction of the arrows changes with respect to the background field.

In Figure~\ref{j2_xyB_t78-123:fig} the current density squared, $j^2$, produced by the QFP wave train launched at Source B is shown in the $x-y$ plane  (`on disk') at height $z=1.42$ in the modeled AR 11166  at two times $t=78\tau_A$, and $t=123\tau_A$. It is evident that $j^2$ is following the funnel structure formed by the background $V_f$ with overall envelope of the QFP waves is nearly stationary. The associate Ohmic heating dissipation proportional to $j^2$ of the waves is maximal at peak $j^2$ values. { Evidently, the background magnetic  field is current free resulting in dark background in this figure.}

The temporal evolution of the velocity and the perturbed magnetic field components at two locations in the model AR are shown in Figure~\ref{dbvnTt_p2:fig}. Figure~\ref{dbvnTt_p2:fig}a shows the temporal evolution of velocity components and magnetic field perturbation compotes at $(-1.49,-1.26,1.7)$ inside the magnetic `funnel', Figure~\ref{dbvnTt_p2:fig}b shows the temporal evolution at $(1.20,-0.01,3.5)$ in the AR above the poles. The corresponding lead times are due to the arrival times of the perturbations from Source~B to the above location. Note, the large amplitudes of the velocity and magnetic field perturbations at location of QFP wave train propagation in the magnetic funnel (indicated with blue `x' in Figure~\ref{drhVEMB_xyB_t123:fig}a), compared to the location away from the magnetic funnel (indicated with yellow `x' in Figure~\ref{drhVEMB_xyB_t123:fig}a). It is evident that the period of the wave is short in the funnel and the phase relations of the velocity components and magnetic field component perturbations are different at the two points, due to the difference in magnetic field magnitude and the corresponding local fast magnetosonic speed $V_f(x,y,z)$ between the two points. In the point away from the QFP wave propagation location at height $z=3.5$, the magnetic field is much weaker than at $z=1.7$, resulting in smaller $V_f$ and longer period of the propagating fast magnetosonic waves.

\section{Discussion and Conclusions} \label{dc:sec}

Analysis of observations of coronal active regions in EUV observed with NASA SDO/AIA satellite shortly after launch, shows clear evidence of QFP wave trains in multiple events associated with flares and CMEs in a number of past studies. The QFP wave trains in ARs were identified as fast magnetosonic waves with speeds of $\sim$1000 km/s, and periods of several minutes showing rapid damping with distance. The QFP wave properties such as periodicity and energy flux provide clues on energy release in impulsive events, on the magnetic structure of coronal ARs. { Analysis of observations show that QFP} waves carry significant energy flux from the energy release site and dissipate the energy in the vicinity of the AR.  

These general properties of QFP waves were modeled previously with idealized MHD models. We extend  previous models by implementing more realistic AR magnetic field using potential magnetic field extrapolation of AR 11166 { using SDO/HMI magnetogram} in the initial state as a case study and include gravitationally stratified solar coronal atmosphere { with typical coronal temperature}. Using 3D MHD model NLRAT, we simulate the excitation and dissipation of QFP magnetosonic waves excited by periodic velocity pulses that model the effects of quasi-periodic flare pulsation at two separate locations. In the present model we employ the periodic velocity driver as the source of the QFP waves associated with flares, while other excitation processes associated with CME's were proposed in the past \citep{Liu12,Mia19}. The damping of the QFP wave occurs primarily due to the spread of the wave energy flux over larger area as the waves propagates away from the small scale source resulting in rapid expansion of the wavefront, accompanied by { enhanced} resistive dissipation of the currents, as well as small numerical viscosity. The { propagating}  waves slow down due to the decreases of the fast magnetosonic speed away from the source. The rapid damping of the QFP waves is consistent with observations.

It is evident that the excitation location near the center of the modeled AR 11166 (Source A) produces waves with weaker directionality, and nearly-circular wave fronts, compared to the source location in the magnetic funnel of the AR, where the directionality and spatial confinement of the QFP wave trains is { significant}, in better agreement with SDO/AIA EUV observations in 171\AA\ and 193\AA. { The directionality of the fast mode wave propagation} is due to the structure of the background fast magnetosonic speed with gradual variability, that acts as a leaky and diffuse waveguide of the QFP waves { in the modeled AR}. 

We find that the 3D MHD model reproduces the main observed { features} of the QFP waves in good qualitative agreement with the QFP waves observed in AR11166 on March 10, 2011 in EUV by SDO/AIA. The present more realistic (compared to previous models) 3D MHD model of the time-dependent EM evolution and { the resulting} details { of the model} provides better understanding of the observed QFP wave generation, propagation and dissipation properties.  The improved 3D MHD modeling results provide a validation tool of coronal seismology diagnostics with QFP waves of AR magnetic structure, and quasi-periodic flare energy release associated with these waves.


The authors acknowledge support by NASA Grant 80NSSC21K1687{, 80NSSC22K0755. The work of LO and TW was also supported by }
Cooperative Agreement 80NSSC21M0180 to Catholic University of America, Partnership for Heliophysics
and Space Environment Research (PHaSER). Resources supporting this work were provided by the NASA High-End Computing (HEC) Program through the NASA Advanced Supercomputing (NAS) Division at Ames Research Center. M.J. acknowledges support from NASA’s SDO/AIA contract (NNG04EA00C) to LMSAL.

%

\vspace{5mm}
\facilities{SDO/AIA}
\software{SolarSoft}


\begin{thebibliography}{}
\expandafter\ifx\csname natexlab\endcsname\relax\def\natexlab#1{#1}\fi
\providecommand{\url}[1]{\href{#1}{#1}}
\providecommand{\dodoi}[1]{doi:~\href{http://doi.org/#1}{\nolinkurl{#1}}}
\providecommand{\doeprint}[1]{\href{http://ascl.net/#1}{\nolinkurl{http://ascl.net/#1}}}
\providecommand{\doarXiv}[1]{\href{https://arxiv.org/abs/#1}{\nolinkurl{https://arxiv.org/abs/#1}}}

\bibitem[{I. {Ballai} {et~al.}(2005){Ballai}, {Erd{\'e}lyi}, \&
  {Pint{\'e}r}}]{Bal05}
{Ballai}, I., {Erd{\'e}lyi}, R., \& {Pint{\'e}r}, B. 2005, \bibinfo{title}{{On
  the Nature of Coronal EIT Waves},} \apjl, 633, L145, \dodoi{10.1086/498447}

\bibitem[{D. {Banerjee} {et~al.}(2021){Banerjee}, {Krishna Prasad}, {Pant},
  {McLaughlin}, {Antolin}, {Magyar}, {Ofman}, {Tian}, {Van Doorsselaere}, {De
  Moortel}, \& {Wang}}]{Ban21}
{Banerjee}, D., {Krishna Prasad}, S., {Pant}, V., {et~al.} 2021,
  \bibinfo{title}{{Magnetohydrodynamic Waves in Open Coronal Structures},}
  \ssr, 217, 76, \dodoi{10.1007/s11214-021-00849-0}

\bibitem[{M.~C.~M. {Cheung} {et~al.}(2015){Cheung}, {Boerner}, {Schrijver},
  {Testa}, {Chen}, {Peter}, \& {Malanushenko}}]{Che15}
{Cheung}, M.~C.~M., {Boerner}, P., {Schrijver}, C.~J., {et~al.} 2015,
  \bibinfo{title}{{Thermal Diagnostics with the Atmospheric Imaging Assembly on
  board the Solar Dynamics Observatory: A Validated Method for Differential
  Emission Measure Inversions},} \apj, 807, 143,
  \dodoi{10.1088/0004-637X/807/2/143}

\bibitem[{K.~P. {Dere}(2020){Dere}}]{Der20}
{Dere}, K.~P. 2020, \bibinfo{title}{{Quiet Sun electron densities and their
  uncertainties derived from spectral emission line intensities},} \mnras, 496,
  2334, \dodoi{10.1093/mnras/staa1645}

\bibitem[{C.~R. {Goddard} {et~al.}(2019){Goddard}, {Nakariakov}, \&
  {Pascoe}}]{God19}
{Goddard}, C.~R., {Nakariakov}, V.~M., \& {Pascoe}, D.~J. 2019,
  \bibinfo{title}{{Fast magnetoacoustic wave trains with time-dependent
  drivers},} \aap, 624, L4, \dodoi{10.1051/0004-6361/201935401}

\bibitem[{C.~R. {Goddard} {et~al.}(2016){Goddard}, {Nistic{\`o}}, {Nakariakov},
  {Zimovets}, \& {White}}]{God16}
{Goddard}, C.~R., {Nistic{\`o}}, G., {Nakariakov}, V.~M., {Zimovets}, I.~V., \&
  {White}, S.~M. 2016, \bibinfo{title}{{Observation of quasi-periodic solar
  radio bursts associated with propagating fast-mode waves},} \aap, 594, A96,
  \dodoi{10.1051/0004-6361/201628478}

\bibitem[{Y. {Guo} {et~al.}(2015){Guo}, {Erd{\'e}lyi}, {Srivastava}, {Hao},
  {Cheng}, {Chen}, {Ding}, \& {Dwivedi}}]{Guo15}
{Guo}, Y., {Erd{\'e}lyi}, R., {Srivastava}, A.~K., {et~al.} 2015,
  \bibinfo{title}{{Magnetohydrodynamic Seismology of a Coronal Loop System by
  the First Two Modes of Standing Kink Waves},} \apj, 799, 151,
  \dodoi{10.1088/0004-637X/799/2/151}

\bibitem[{J. {Hu} {et~al.}(2024){Hu}, {Ye}, {Chen}, {Mei}, {Tang}, \&
  {Lin}}]{Hu24}
{Hu}, J., {Ye}, J., {Chen}, Y., {et~al.} 2024, \bibinfo{title}{{Excitation of
  Quasiperiodic Fast-propagating Waves in the Early Stage of the Solar
  Eruption},} \apj, 962, 42, \dodoi{10.3847/1538-4357/ad1993}

\bibitem[{D.~Y. {Kolotkov} {et~al.}(2021){Kolotkov}, {Nakariakov}, {Moss}, \&
  {Shellard}}]{Kol21}
{Kolotkov}, D.~Y., {Nakariakov}, V.~M., {Moss}, G., \& {Shellard}, P. 2021,
  \bibinfo{title}{{Fast magnetoacoustic wave trains: from tadpoles to
  boomerangs},} \mnras, 505, 3505, \dodoi{10.1093/mnras/stab1587}

\bibitem[{J.~R. {Lemen} {et~al.}(2012){Lemen}, {Title}, {Akin}, {Boerner},
  {Chou}, {Drake}, {Duncan}, {Edwards}, {Friedlaender}, {Heyman}, {Hurlburt},
  {Katz}, {Kushner}, {Levay}, {Lindgren}, {Mathur}, {McFeaters}, {Mitchell},
  {Rehse}, {Schrijver}, {Springer}, {Stern}, {Tarbell}, {Wuelser}, {Wolfson},
  {Yanari}, {Bookbinder}, {Cheimets}, {Caldwell}, {Deluca}, {Gates}, {Golub},
  {Park}, {Podgorski}, {Bush}, {Scherrer}, {Gummin}, {Smith}, {Auker},
  {Jerram}, {Pool}, {Soufli}, {Windt}, {Beardsley}, {Clapp}, {Lang}, \&
  {Waltham}}]{Lem12}
{Lemen}, J.~R., {Title}, A.~M., {Akin}, D.~J., {et~al.} 2012,
  \bibinfo{title}{{The Atmospheric Imaging Assembly (AIA) on the Solar Dynamics
  Observatory (SDO)},} \solphys, 275, 17, \dodoi{10.1007/s11207-011-9776-8}

\bibitem[{L. {Li} {et~al.}(2018){Li}, {Zhang}, {Peter}, {Chitta}, {Su}, {Song},
  {Xia}, \& {Hou}}]{Li18a}
{Li}, L., {Zhang}, J., {Peter}, H., {et~al.} 2018,
  \bibinfo{title}{{Quasi-periodic Fast Propagating Magnetoacoustic Waves during
  the Magnetic Reconnection Between Solar Coronal Loops},} \apjl, 868, L33,
  \dodoi{10.3847/2041-8213/aaf167}

\bibitem[{W. {Liu} \& L. {Ofman}(2014){Liu} \& {Ofman}}]{LO14}
{Liu}, W., \& {Ofman}, L. 2014, \bibinfo{title}{{Advances in Observing Various
  Coronal EUV Waves in the SDO Era and Their Seismological Applications
  (Invited Review)},} \solphys, 289, 3233, \dodoi{10.1007/s11207-014-0528-4}

\bibitem[{W. {Liu} {et~al.}(2012){Liu}, {Ofman}, {Nitta}, {Aschwanden},
  {Schrijver}, {Title}, \& {Tarbell}}]{Liu12}
{Liu}, W., {Ofman}, L., {Nitta}, N.~V., {et~al.} 2012,
  \bibinfo{title}{{Quasi-periodic Fast-mode Wave Trains within a Global EUV
  Wave and Sequential Transverse Oscillations Detected by SDO/AIA},} \apj, 753,
  52, \dodoi{10.1088/0004-637X/753/1/52}

\bibitem[{W. {Liu} {et~al.}(2011){Liu}, {Title}, {Zhao}, {Ofman}, {Schrijver},
  {Aschwanden}, {De Pontieu}, \& {Tarbell}}]{Liu11}
{Liu}, W., {Title}, A.~M., {Zhao}, J., {et~al.} 2011, \bibinfo{title}{{Direct
  Imaging of Quasi-periodic Fast Propagating Waves of \~{}2000 km s$^{-1}$ in
  the Low Solar Corona by the Solar Dynamics Observatory Atmospheric Imaging
  Assembly},} \apjl, 736, L13, \dodoi{10.1088/2041-8205/736/1/L13}

\bibitem[{Y. {Miao} {et~al.}(2021){Miao}, {Li}, {Yuan}, {Jiang}, {Elmhamdi},
  {Zhao}, \& {Anfinogentov}}]{Mia21}
{Miao}, Y., {Li}, D., {Yuan}, D., {et~al.} 2021, \bibinfo{title}{{Diagnosing a
  Solar Flaring Core with Bidirectional Quasi-periodic Fast Propagating
  Magnetoacoustic Waves},} \apjl, 908, L37, \dodoi{10.3847/2041-8213/abdfce}

\bibitem[{Y. {Miao} {et~al.}(2020){Miao}, {Liu}, {Elmhamdi}, {Kordi}, {Shen},
  {Al-Shammari}, {Al-Mosabeh}, {Jiang}, \& {Yuan}}]{Mia20}
{Miao}, Y., {Liu}, Y., {Elmhamdi}, A., {et~al.} 2020, \bibinfo{title}{{Two
  Quasi-periodic Fast-propagating Magnetosonic Wave Events Observed in Active
  Region NOAA 11167},} \apj, 889, 139, \dodoi{10.3847/1538-4357/ab655f}

\bibitem[{Y.~H. {Miao} {et~al.}(2019){Miao}, {Liu}, {Shen}, {Li}, {Abidin},
  {Elmhamdi}, \& {Kordi}}]{Mia19}
{Miao}, Y.~H., {Liu}, Y., {Shen}, Y.~D., {et~al.} 2019, \bibinfo{title}{{A
  Quasi-periodic Propagating Wave and Extreme-ultraviolet Waves Excited
  Simultaneously in a Solar Eruption Event},} \apjl, 871, L2,
  \dodoi{10.3847/2041-8213/aafaf9}

\bibitem[{S. {Mondal} {et~al.}(2024){Mondal}, {Srivastava}, {Pontin}, {Priest},
  {Kwon}, \& {Yuan}}]{Mon24}
{Mondal}, S., {Srivastava}, A.~K., {Pontin}, D.~I., {et~al.} 2024,
  \bibinfo{title}{{Generation of Fast Magnetoacoustic Waves in the Corona by
  Impulsive Bursty Reconnection},} \apj, 977, 235,
  \dodoi{10.3847/1538-4357/ad9022}

\bibitem[{K. {Murawski} {et~al.}(2001){Murawski}, {Nakariakov}, \&
  {Pelinovsky}}]{Mur01}
{Murawski}, K., {Nakariakov}, V.~M., \& {Pelinovsky}, E.~N. 2001,
  \bibinfo{title}{{Fast magnetoacoustic waves in a randomly structured solar
  corona},} \aap, 366, 306, \dodoi{10.1051/0004-6361:20000027}

\bibitem[{V.~M. Nakariakov \& D.~Y. Kolotkov(2020)Nakariakov \&
  Kolotkov}]{NK20}
Nakariakov, V.~M., \& Kolotkov, D.~Y. 2020, \bibinfo{title}{Magnetohydrodynamic
  Waves in the Solar Corona,} Annual Review of Astronomy and Astrophysics, 58,
  441, \dodoi{https://doi.org/10.1146/annurev-astro-032320-042940}

\bibitem[{V.~M. {Nakariakov} {et~al.}(2024){Nakariakov}, {Zhong}, {Kolotkov},
  {Meadowcroft}, {Zhong}, \& {Yuan}}]{Nak24}
{Nakariakov}, V.~M., {Zhong}, S., {Kolotkov}, D.~Y., {et~al.} 2024,
  \bibinfo{title}{{Diagnostics of the solar coronal plasmas by
  magnetohydrodynamic waves: magnetohydrodynamic seismology},} Reviews of
  Modern Plasma Physics, 8, 19, \dodoi{10.1007/s41614-024-00160-9}

\bibitem[{G. {Nistic{\`o}} {et~al.}(2014){Nistic{\`o}}, {Pascoe}, \&
  {Nakariakov}}]{Nis14}
{Nistic{\`o}}, G., {Pascoe}, D.~J., \& {Nakariakov}, V.~M. 2014,
  \bibinfo{title}{{Observation of a high-quality quasi-periodic rapidly
  propagating wave train using SDO/AIA},} \aap, 569, A12,
  \dodoi{10.1051/0004-6361/201423763}

\bibitem[{L. {Ofman} \& J.~M. {Davila}(1995){Ofman} \& {Davila}}]{OD95}
{Ofman}, L., \& {Davila}, J.~M. 1995, \bibinfo{title}{{Alfv{\'e}n wave heating
  of coronal holes and the relation to the high-speed solar wind},} J. Geophys.
  Res., 100, 23413

\bibitem[{L. {Ofman} \& W. {Liu}(2018){Ofman} \& {Liu}}]{OL18}
{Ofman}, L., \& {Liu}, W. 2018, \bibinfo{title}{{Quasi-periodic
  Counter-propagating Fast Magnetosonic Wave Trains from Neighboring Flares:
  SDO/AIA Observations and 3D MHD Modeling},} \apj, 860, 54,
  \dodoi{10.3847/1538-4357/aac2e8}

\bibitem[{L. {Ofman} {et~al.}(2011){Ofman}, {Liu}, {Title}, \&
  {Aschwanden}}]{Ofm11}
{Ofman}, L., {Liu}, W., {Title}, A., \& {Aschwanden}, M. 2011,
  \bibinfo{title}{{Modeling Super-fast Magnetosonic Waves Observed by SDO in
  Active Region Funnels},} \apjl, 740, L33, \dodoi{10.1088/2041-8205/740/2/L33}

\bibitem[{L. {Ofman} {et~al.}(2015){Ofman}, {Parisi}, \& {Srivastava}}]{Ofm15b}
{Ofman}, L., {Parisi}, M., \& {Srivastava}, A.~K. 2015,
  \bibinfo{title}{{Three-dimensional MHD modeling of vertical kink oscillations
  in an active region plasma curtain},} \aap, 582, A75,
  \dodoi{10.1051/0004-6361/201425054}

\bibitem[{L. {Ofman} \& B.~J. {Thompson}(2002){Ofman} \& {Thompson}}]{OT02}
{Ofman}, L., \& {Thompson}, B.~J. 2002, \bibinfo{title}{{Interaction of EIT
  Waves with Coronal Active Regions},} \apj, 574, 440, \dodoi{10.1086/340924}

\bibitem[{L. {Ofman} \& T. {Wang}(2022){Ofman} \& {Wang}}]{OW22}
{Ofman}, L., \& {Wang}, T. 2022, \bibinfo{title}{{Excitation and Damping of
  Slow Magnetosonic Waves in Flaring Hot Coronal Loops: Effects of Compressive
  Viscosity},} \apj, 926, 64, \dodoi{10.3847/1538-4357/ac4090}

\bibitem[{L. {Ofman} {et~al.}(2012){Ofman}, {Wang}, \& {Davila}}]{Ofm12}
{Ofman}, L., {Wang}, T.~J., \& {Davila}, J.~M. 2012, \bibinfo{title}{{Slow
  Magnetosonic Waves and Fast Flows in Active Region Loops},} \apj, 754, 111,
  \dodoi{10.1088/0004-637X/754/2/111}

\bibitem[{D.~J. {Pascoe} {et~al.}(2017){Pascoe}, {Goddard}, \&
  {Nakariakov}}]{Pas17}
{Pascoe}, D.~J., {Goddard}, C.~R., \& {Nakariakov}, V.~M. 2017,
  \bibinfo{title}{{Dispersive Evolution of Nonlinear Fast Magnetoacoustic Wave
  Trains},} \apjl, 847, L21, \dodoi{10.3847/2041-8213/aa8db8}

\bibitem[{D.~J. {Pascoe} {et~al.}(2013){Pascoe}, {Nakariakov}, \&
  {Kupriyanova}}]{Pas13}
{Pascoe}, D.~J., {Nakariakov}, V.~M., \& {Kupriyanova}, E.~G. 2013,
  \bibinfo{title}{{Fast magnetoacoustic wave trains in magnetic funnels of the
  solar corona},} \aap, 560, A97, \dodoi{10.1051/0004-6361/201322678}

\bibitem[{D.~J. {Pascoe} {et~al.}(2014){Pascoe}, {Nakariakov}, \&
  {Kupriyanova}}]{Pas14}
{Pascoe}, D.~J., {Nakariakov}, V.~M., \& {Kupriyanova}, E.~G. 2014,
  \bibinfo{title}{{Fast magnetoacoustic wave trains in coronal holes},} \aap,
  568, A20, \dodoi{10.1051/0004-6361/201423931}

\bibitem[{W.~D. {Pesnell} {et~al.}(2012){Pesnell}, {Thompson}, \&
  {Chamberlin}}]{Pes12}
{Pesnell}, W.~D., {Thompson}, B.~J., \& {Chamberlin}, P.~C. 2012,
  \bibinfo{title}{{The Solar Dynamics Observatory (SDO)},} \solphys, 275, 3,
  \dodoi{10.1007/s11207-011-9841-3}

\bibitem[{E. {Provornikova} {et~al.}(2018){Provornikova}, {Ofman}, \&
  {Wang}}]{POW18}
{Provornikova}, E., {Ofman}, L., \& {Wang}. 2018, \bibinfo{title}{Excitation of
  flare-induced waves in coronal loops and the effects of radiative cooling,}
  AdSpR, 61, 645 , \dodoi{10.1016/j.asr.2017.07.042}

\bibitem[{T. {Sakurai}(1989){Sakurai}}]{Sak89}
{Sakurai}, T. 1989, \bibinfo{title}{{Computational modeling of magnetic fields
  in solar active regions},} \ssr, 51, 11, \dodoi{10.1007/BF00226267}

\bibitem[{Y. {Shen} {et~al.}(2018){Shen}, {Liu}, {Song}, \& {Tian}}]{She18}
{Shen}, Y., {Liu}, Y., {Song}, T., \& {Tian}, Z. 2018, \bibinfo{title}{{A
  Quasi-periodic Fast-propagating Magnetosonic Wave Associated with the
  Eruption of a Magnetic Flux Rope},} \apj, 853, 1,
  \dodoi{10.3847/1538-4357/aaa3ff}

\bibitem[{Y. {Shen} {et~al.}(2017){Shen}, {Liu}, {Tian}, \& {Qu}}]{She17}
{Shen}, Y., {Liu}, Y., {Tian}, Z., \& {Qu}, Z. 2017, \bibinfo{title}{{On a
  Small-scale EUV Wave: The Driving Mechanism and the Associated Oscillating
  Filament},} \apj, 851, 101, \dodoi{10.3847/1538-4357/aa9af0}

\bibitem[{Y. {Shen} {et~al.}(2022){Shen}, {Zhou}, {Duan}, {Tang}, {Zhou}, \&
  {Tan}}]{She22}
{Shen}, Y., {Zhou}, X., {Duan}, Y., {et~al.} 2022, \bibinfo{title}{{Coronal
  Quasi-periodic Fast-mode Propagating Wave Trains},} \solphys, 297, 20,
  \dodoi{10.1007/s11207-022-01953-2}

\bibitem[{Y.-D. {Shen} \& Y. {Liu}(2012){Shen} \& {Liu}}]{She12}
{Shen}, Y.-D., \& {Liu}, Y. 2012, \bibinfo{title}{{Observational Study of the
  Quasi-periodic Fast-propagating Magnetosonic Waves and the Associated Flare
  on 2011 May 30},} \apj, 753, 53, \dodoi{10.1088/0004-637X/753/1/53}

\bibitem[{Y.-D. {Shen} {et~al.}(2013){Shen}, {Liu}, {Su}, {Li}, {Zhang},
  {Tian}, {Zhao}, \& {Elmhamdi}}]{She13}
{Shen}, Y.-D., {Liu}, Y., {Su}, J.-T., {et~al.} 2013,
  \bibinfo{title}{{Observations of a Quasi-periodic, Fast-Propagating
  Magnetosonic Wave in Multiple Wavelengths and Its Interaction with Other
  Magnetic Structures},} \solphys, 288, 585, \dodoi{10.1007/s11207-013-0395-4}

\bibitem[{A.~K. {Srivastava} {et~al.}(2025){Srivastava}, {Mondal}, {Priest},
  {Mishra}, {Pontin}, {Kwon}, {Yuan}, {Murawski}, \& {Asai}}]{Sri25}
{Srivastava}, A.~K., {Mondal}, S., {Priest}, E.~R., {et~al.} 2025,
  \bibinfo{title}{{Localized Heating and Dynamics of the Solar Corona due to a
  Symbiosis of Waves and Reconnection},} \apj, 984, 36,
  \dodoi{10.3847/1538-4357/adc379}

\bibitem[{X. {Sun}(2013){Sun}}]{Sun13}
{Sun}, X. 2013, \bibinfo{title}{{On the Coordinate System of Space-Weather HMI
  Active Region Patches (SHARPs): A Technical Note},} arXiv e-prints,
  arXiv:1309.2392, \dodoi{10.48550/arXiv.1309.2392}

\bibitem[{S. {Takasao} \& K. {Shibata}(2016){Takasao} \& {Shibata}}]{TS16}
{Takasao}, S., \& {Shibata}, K. 2016, \bibinfo{title}{{Above-the-loop-top
  Oscillation and Quasi-periodic Coronal Wave Generation in Solar Flares},}
  \apj, 823, 150, \dodoi{10.3847/0004-637X/823/2/150}

\bibitem[{T. {Van Doorsselaere} {et~al.}(2016){Van Doorsselaere},
  {Kupriyanova}, \& {Yuan}}]{Van16}
{Van Doorsselaere}, T., {Kupriyanova}, E.~G., \& {Yuan}, D. 2016,
  \bibinfo{title}{{Quasi-periodic Pulsations in Solar and Stellar Flares: An
  Overview of Recent Results (Invited Review)},} \solphys, 291, 3143,
  \dodoi{10.1007/s11207-016-0977-z}

\bibitem[{T. {Wang} {et~al.}(2024){Wang}, {Ofman}, \& {Bradshaw}}]{wang24}
{Wang}, T., {Ofman}, L., \& {Bradshaw}, S.~J. 2024, \bibinfo{title}{{Exploring
  Standing and Reflected Slow-Mode Waves in Flaring Coronal Loops: A Parametric
  Study Using 2.5D MHD Modeling},} \solphys, 299, 37,
  \dodoi{10.1007/s11207-024-02285-z}

\bibitem[{T. {Wang} {et~al.}(2021){Wang}, {Ofman}, {Yuan}, {Reale}, {Kolotkov},
  \& {Srivastava}}]{Wan21}
{Wang}, T., {Ofman}, L., {Yuan}, D., {et~al.} 2021, \bibinfo{title}{{Slow-Mode
  Magnetoacoustic Waves in Coronal Loops},} \ssr, 217, 34,
  \dodoi{10.1007/s11214-021-00811-0}

\bibitem[{T.~J. {Wang}(2016){Wang}}]{Wan16}
{Wang}, T.~J. 2016, \bibinfo{title}{{Waves in Solar Coronal Loops},} Washington
  DC American Geophysical Union Geophysical Monograph Series, 216, 395,
  \dodoi{10.1002/9781119055006.ch23}

\bibitem[{G.~L. {Withbroe} \& R.~W. {Noyes}(1977){Withbroe} \& {Noyes}}]{WN77}
{Withbroe}, G.~L., \& {Noyes}, R.~W. 1977, \bibinfo{title}{{Mass and energy
  flow in the solar chromosphere and corona},} \araa, 15, 363,
  \dodoi{10.1146/annurev.aa.15.090177.002051}

\bibitem[{D. {Yuan} {et~al.}(2013){Yuan}, {Shen}, {Liu}, {Nakariakov}, {Tan},
  \& {Huang}}]{Yua13}
{Yuan}, D., {Shen}, Y.-D., {Liu}, Y., {et~al.} 2013, \bibinfo{title}{{Distinct
  propagating fast wave trains associated with flaring energy releases},} \aap,
  554, A144, \dodoi{10.1051/0004-6361/201321435}

\bibitem[{Y. {Zhang} {et~al.}(2015){Zhang}, {Zhang}, {Wang}, \&
  {Nakariakov}}]{Zha15}
{Zhang}, Y., {Zhang}, J., {Wang}, J., \& {Nakariakov}, V.~M. 2015,
  \bibinfo{title}{{Coexisting fast and slow propagating waves of the extreme-UV
  intensity in solar coronal plasma structures},} \aap, 581, A78,
  \dodoi{10.1051/0004-6361/201525621}

\bibitem[{X. {Zhou} {et~al.}(2022){Zhou}, {Shen}, {Liang}, {Qu}, {Duan},
  {Tang}, {Zhou}, \& {Tan}}]{Zho22}
{Zhou}, X., {Shen}, Y., {Liang}, H., {et~al.} 2022, \bibinfo{title}{{Recurrent
  Narrow Quasiperiodic Fast-propagating Wave Trains Excited by the Intermittent
  Energy Release in the Accompanying Solar Flare},} \apj, 941, 59,
  \dodoi{10.3847/1538-4357/aca1b6}

\bibitem[{X. {Zhou} {et~al.}(2024){Zhou}, {Tang}, {Qu}, {Yu}, {Zhou}, {Xiang},
  {Ibrahim}, \& {Shen}}]{Zho24}
{Zhou}, X., {Tang}, Z., {Qu}, Z., {et~al.} 2024, \bibinfo{title}{{On the Origin
  of a Broad Quasiperiodic Fast-propagating Wave Train: Unwinding Jet as the
  Driver},} \apjl, 974, L3, \dodoi{10.3847/2041-8213/ad7a68}

\end{thebibliography}
\end{document}